\documentclass{article}
\usepackage[utf8]{inputenc}
\usepackage{authblk}
\usepackage{setspace}
\usepackage[margin=1.25in]{geometry}
\usepackage{graphicx}
\graphicspath{ {./figures/} }
\usepackage{subcaption}
\usepackage{amsmath}
\usepackage{lineno}
\usepackage{booktabs}
\usepackage{hyperref}
\usepackage{newunicodechar}
\newunicodechar{ }{~}

\usepackage{array}
\usepackage{graphicx}
\usepackage{tikz}
\usetikzlibrary{positioning,calc,decorations.pathreplacing,shadows.blur}

\usepackage[style=numeric-comp, 
citestyle=numeric-comp,
sorting=none]{biblatex}
\usepackage{amssymb}
\title{\textbf{Automated MRI Tumor Segmentation using hybrid U-Net with Transformer and Efficient Attention}}

\author[1*]{\textbf{Syed Haider Ali}}
\author[2]{\textbf{Asrar Ahmad}}
\author[1,2]{\textbf{Muhammad Ali}}
\author[3,4,5]{\textbf{Asifullah Khan}}
\author[1,3]{\textbf{Nadeem Shaukat}}

\affil[1]{Department of Physics \& Applied Mathematics, Pakistan Institute of Engineering and Applied Sciences (PIEAS), P. O. Nilore 45650, Islamabad .}
\affil[2]{Department of Medical Sciences, Pakistan Institute of Engineering and Applied Sciences (PIEAS), P. O. Nilore 45650, Islamabad .}
\affil[3]{Center for Mathematical Sciences, Pakistan Institute of Engineering and Applied Sciences (PIEAS), P. O. Nilore 45650, Islamabad .}
\affil[4]{Pattern Recognition Lab, Department of Computer \& Information Sciences, Pakistan Institute of Engineering \& Applied Sciences, Nilore, Islamabad, 45650, Pakistan }
\affil[5]{PIEAS Artificial Intelligence Center (PAIC), Pakistan Institute of Engineering \& Applied Sciences, Nilore, Islamabad, 45650, Pakistan }
\affil[*]{Address correspondence to: syedhaider.ali2021453@gmail.com}

\begin{document}

\maketitle
\newpage
\begin{abstract}
Cancer is an abnormal growth with the potential to invade locally and metastasize to distant organs. Auto-segmentation of the tumor and surrounding normal tissues is required for radiotherapy treatment plan optimization. Recent studies have proposed various AI based segmentation models, mostly trained on massively online available datasets. These datasets lack the heterogeneity of local patient populations, and while such studies do advance the research in AI based medical image segmentation, studies on local datasets need to be prioritized if the future aim is to develop and integrate AI based tumor segmentation models directly in the hospitals softwares for efficient and accurate oncology treatment planning and execution. This study aims to enhance tumor segmentation using computationally efficient and accurate UNET-Transformer hybrid models on magnetic resonance imaging (MRI) datasets that were taken from a local hospital while taking care of the patient privacy. A robust data pipeline was developed for seamless extraction and preprocessing of DICOM files, followed by extensive image augmentation to ensure model generalization in diverse clinical settings bringing a total dataset size of 6080 images for training. A novel approach is proposed that integrates UNET-based convolutional neural networks (CNNs) with a transformer bottleneck and complementary attention modules, including efficient attention, Squeeze-and-Excitation (SE) blocks, Convolutional Block Attention Module (CBAM), and ResNeXt blocks. To accelerate convergence and reduce computational demands, a maximum batchsize of 8 images was chosen and the encoder was loaded with pretrained weights from ImageNet to conserve computational resources since the model was trained on the T4 x2 GPUs offered by Kaggle. Given the restricted runtime of kaggle’s GPU kernel, a checkpointing system was implemented to resume uninterrupted training progressing across multiple sessions, facilitating efficient utilization of available computational resources. Quantitative evaluation on the local MRI dataset yielded a Dice similarity coefficient of 76.4\% and an Intersection over Union (IoU) of 73.6\%, demonstrating competitive performance despite the relatively limited dataset size.

\end{abstract}

\newpage
\section{INTRODUCTION}
Improper classification and identification of brain tumors is a major cause of high mortality rate worldwide in afflicted patients. An estimated 4,407 deaths from brain and central nervous system cancers occurred in Pakistan in 2022, highlighting the critical need for improved diagnosis and treatment methods \cite{4}. Magnetic resonance imaging (MRI) offers superior soft-tissue contrast, making it the modality of choice for identifying brain tumors and lesions \cite{Essig2012MR}. Precise segmentation of these abnormalities is vital for both early detection and the formulation of appropriate treatment strategies. An ischemic stroke results from the obstruction of a cerebral blood vessel, which deprives brain tissue of essential oxygen and nutrients. Such an event represents a critical medical emergency that may lead to irreversible neurological damage or fatality\cite{Adams1993Classification}. Manual segmentation remains the clinical gold standard for delineating tumor boundaries, despite being time-consuming and subject to inter-observer variability\cite{Warfield2004STAPLE}. Medical experts in local hospitals still use manual segmentation of tumor regions to mark areas and boundaries of damaged brain tissue.
\paragraph{}
Manual segmentation has its own daunting challenges which impact the patient directly. It is a time-consuming process and often a difficult task left out for the oncologists. Unlike well-defined organs (eyes, brainstem), tumors are irregular and unpredictable. They have inconsistent shape, size, boundaries and often blend with surrounding tissues, making detection harder. Dynamic growth and irregular borders complicate precise segmentation, and the contouring is often subjective to the view of the oncologist. There is a need for automated methods for tumor segmentation that can offer a more efficient and automated approach to diagnosing and treating tumors\cite{Litjens2017Survey}. Supervised deep learning methods like CNNs have contributed to image segmentation tasks in medical physics greatly over the years, especially reflecting in the BRATS\cite{menze2015multimodal} challenges annually. UNet \cite{unet} is one of the most used deep learning architecture in medical image segmentation. The UNET architecture consists of an encoder and decoder block linked with a bottleneck layer in a U shape (hence the name UNet). The purpose of the encoder block is to downsample the input image using convolutional and pooling layers to condense the information in it by contracting the spatial resolution and depth structures, passing it to a bottleneck layer that connects it to the decoder layer. The decoder layer then upsamples the condensed image to an output label mask using Transpose Convolutions.  
\paragraph{}
While pure CNNs excel at capturing local features, they struggle with long-range dependencies. A recent survey of self-supervised learning mechanisms for Vision Transformers demonstrates how SSL-pretrained ViTs can greatly reduce labeled data requirements while preserving downstream segmentation accuracy \cite{survey_ssl_vit_2408.17059}. At the same time, an extensive review of Vision Transformer–based models for medical image segmentation reports that hybrid CNN–ViT architectures consistently outperform pure CNNs on MRI and CT benchmarks by better modeling global context \cite{survey_medseg_vit_2312.00634}.  
\paragraph{}
While models like UNET and its variants have shown remarkable success in the BRATS challenges, they face significant hurdles when applied to real-world clinical data, particularly in resource-limited settings. The reliance on curated datasets such as BRATS poses critical challenges for models aimed to be deployed locally in hospitals for radiation oncology treatment planning. Public datasets often fail to capture the heterogeneity of local patient populations, imaging protocols, and scanner variations. In addition, Models trained on BRATS or similar datasets may not generalize well to localized datasets, which are smaller in size and exhibit greater variability\cite{Guan2022Domain}. Data privacy concerns and regulations restrict the availability of medical data, thus limiting the prospect of such studies to test out models on vast amount of data\cite{GDPR2016}. 

To address this gap,  a novel hybrid UNET-Transformer\cite{3} architecture is proposed and trained and evaluated on a localized hospital dataset. This dataset reflects the real-world clinical heterogeneity in patient populations, imaging modalities, and annotation practices. the present approach emphasizes the feasibility and challenges of deploying deep learning models in resource-constrained settings, where public datasets may not fully capture local variability. 
\paragraph{}
 This study highlights the importance of tailoring automated solutions to local clinical while also providing insights into the translational challenges of applying state-of-the-art deep learning techniques in under-resourced settings. This approach represents a \textbf{novel contribution} introducing hybrid image segmentation models designed to handle the heterogeneity of local datasets, while comparing its performance against models trained on large open data to highlight the generalization gap and practical challenges in clinical deployment.

\section{Materials and Methods}
\subsection{Dataset and Data Preparation}

The dataset used in this study was directly collected from a local hospital, offering a unique opportunity to evaluate our segmentation method on authentic clinical data. Unlike standardized benchmarks such as the BraTS dataset, which is employed in state‐of‐the‐art approaches like nnU-Net \cite{dataset1} and UNETR \cite{dataset2} or the MICCAI LiTS \cite{bilic2019liver} dataset used in H-DenseUNet \cite{HDENSENET}, our dataset reflects the natural variability encountered in routine clinical practice. The inherent differences in imaging protocols and patient demographics introduce challenges for segmentation models, while simultaneously enhancing the real-world relevance of our findings. 

As summarized in Table \ref{tab:mri_dataset}, the dataset comprises MRI scans from six patients, with four cases acquired using T1-weighted sequences and two with T2-weighted sequences. Each patient contributed approximately 175 slices, yielding a total of 700 T1-weighted and 350 T2-weighted images. All images were acquired at a voxel resolution of acquired at a voxel resolution of $0.84\,\times\,0.84\,\times\,2.00 mm$. The raw data, stored as DICOM files, were initially converted into volumetric images using the open-source PyDICOM library \cite{pydicom}. Expert-provided segmentation masks were then transformed into binary label maps, and the volumetric data were sliced into 2D images, each paired with its corresponding mask. These processed images were subsequently organized into TensorFlow data pipelines to facilitate efficient model training and evaluation. 

\paragraph{}

\begin{table}[ht]
\centering
\renewcommand{\arraystretch}{1.5}
\setlength{\tabcolsep}{5pt}
\begin{tabular}{|>{\centering\arraybackslash}m{2cm}|>{\centering\arraybackslash}m{2cm}|>{\centering\arraybackslash}m{2cm}|>{\centering\arraybackslash}m{2cm}|>{\centering\arraybackslash}m{4cm}|}
\hline
\textbf{MRI Modality} & \textbf{Number of patients} & \textbf{Images per patient} & \textbf{Total Images} & \textbf{Voxel Size} \\ \hline
T1 & 4 & $\sim$175 & $\sim$700 & 0.84x0.84x2.00 mm \\ \hline
T2 & 2 & $\sim$175 & $\sim$350 & 0.84x0.84x2.00 mm \\ \hline
\end{tabular}
\caption{Summary of MRI dataset characteristics.}
\label{tab:mri_dataset}
\end{table}

\begin{figure}
    \centering
    \includegraphics[width=0.6\linewidth]{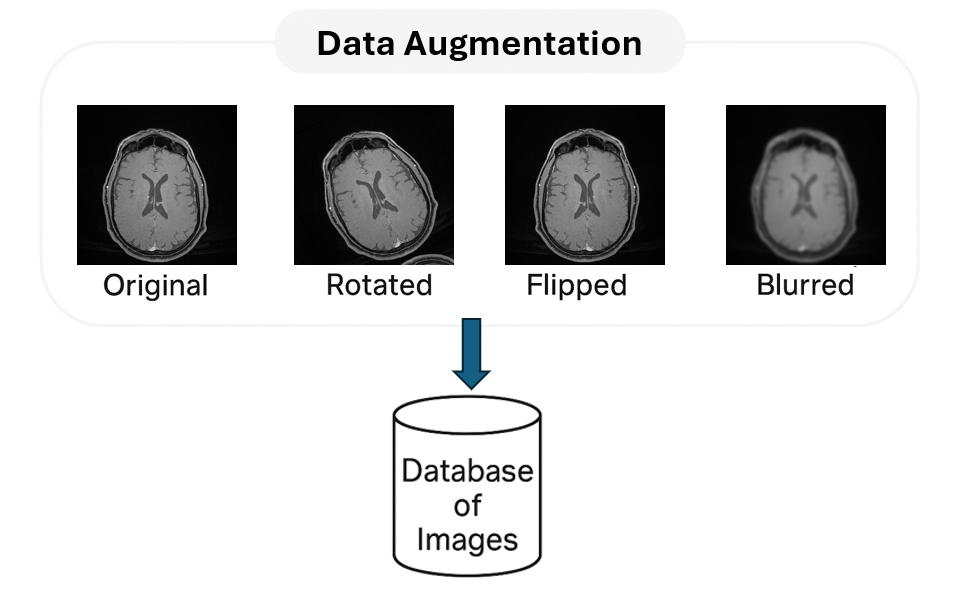}
    \caption{Data Augmentation Pipeline}
    \label{fig:augpipe}
\end{figure}

Given the modest size of the original dataset (\~ 1,000 images), an extensive data augmentation strategy was implemented to enhance both data diversity and model robustness. As illustrated in Figure \ref{fig:augpipe}, several data augmentation techniques including flips, rotations, Gaussian blur, and contrast shifting are applied that expand the size of the dataset to 6,080 images. Such augmentation methods have been shown to improve the generalization and accuracy of deep learning models in medical imaging tasks \cite{dataset4}.

\begin{figure}
    \centering
    \includegraphics[width=0.6\linewidth]{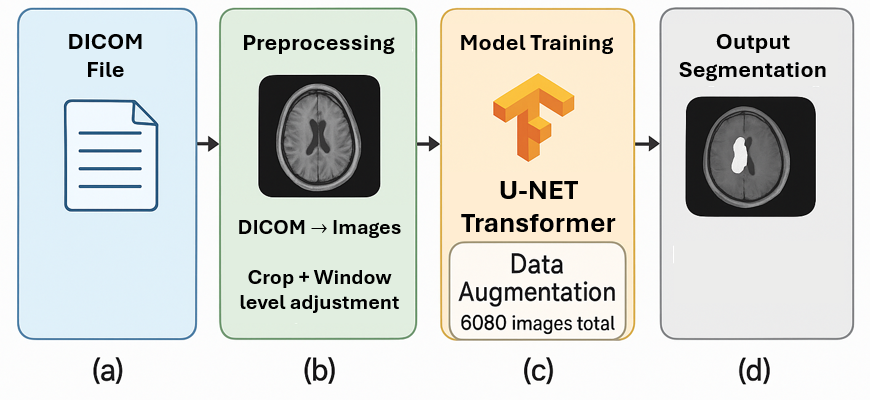}
    \caption{Data Processing pipeline}
    \label{fig: Pipeline}
\end{figure}

Reliance on clinically sourced data is particularly significant, as it captures a wide range of acquisition parameters and patient variability that standardized datasets may not reflect. This approach does, however, raise important ethical considerations. All patient images were rigorously de-identified and handled in accordance with ethical guidelines approved by our local hospital, following best practices for the ethical use of medical imaging data. 

By integrating a carefully designed preprocessing pipeline (see Figure \ref{fig: Pipeline}) with a robust augmentation strategy (see Figure \ref{fig:augpipe}), the dataset effectively bridges the gap between experimental validation and practical, real-world application. The inherent variability of clinical imaging, coupled with the quantitative performance improvements demonstrated by our method, underscores the potential impact of this study on future clinical workflows and personalized treatment planning.

\subsection{Proposed Methodology}
\begin{figure}
    \centering
    \includegraphics[width=0.5\linewidth]{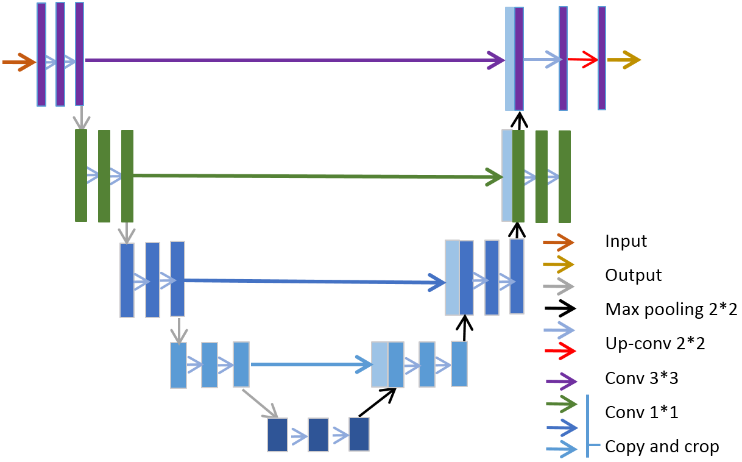}
    \caption{(left) The Encoder block that downsamples the image and passes it through a bottleneck layer.
    (right) The decoder block resizes the tensor into an output segmented image. Adapted from \cite{unetfigure}}
    \label{fig:UNET}
\end{figure}

The UNET Architecture is an appropriate choice when it comes to image segmentation. Segmentation is a special case of image classification, where each pixel in an image is classified, the result in a segmented image corresponding an input image. Convolutional filters in the UNET architecture acts as feature extractors , that extract relevant features from an image such as edges, corners and textures. CNNs process images layer by layer in a hierarchical structure, gradually building up representations of increasingly complex features. Figure \ref{fig:UNET} shows the encoder-decoder structure of a UNET. This structure allows CNNs to capture local patterns effectively.
\paragraph{}
Vaswani et al \cite{attention} introduced the transformer architecture, which employs a self attention mechanism to process sequences concurrently, yielding high efficiency. This approach enables the network to attend to the most pertinent regions of an image by mapping a query against collections of keys and values to produce an output. A compatibility function between the query and each key computes weights for the associated values, which are then aggregated through a weighted sum to form the final output \cite{attention}. By assigning attention weights in this way, the model captures long range dependencies and attains a wider receptive field than convolutional neural networks, thereby integrating broader contextual information. Combining the merits of both transformer and convolutional designs, hybrid models have shown remarkable performance improvement over baseline models.

\begin{equation}
    Attention(Q,K,V) = softmax(\frac{QK^T}{\sqrt{d_k}})V
    \label{eq1}
\end{equation}

Hybrid architectures have emerged as a significant advancement in deep learning, particularly for medical image segmentation tasks\cite{chen2021transunet}. By leveraging the strengths of multiple methodologies, these architectures aim to improve performance and address inherent limitations in conventional models. Traditional UNet architectures, renowned for their inductive bias and efficient handling of small datasets, often struggle to capture long-range dependencies\cite{unet}. Subsequently, transformer-based models excel in modeling global relationships but demand extensive annotated data, making them impractical for resource-constrained environments\cite{dosovitskiy2020image}. Hybrid models offer a balanced solution by integrating the complementary features of these approaches\cite{dataset2}.
\paragraph{}
Several hybrid architectures have been proposed in the literature. The Hybrid UNet Transformer (HUT) \cite{1} combines convolutional layers with transformers, introducing transformers at skip connections to enhance receptive fields. While effective in improving attention mechanisms, its reliance on large annotated datasets limits its applicability to resource-constrained settings. Similarly, the MaxViT-UNet framework \cite{3} integrates MaxViT blocks into an encoder-decoder structure, achieving significant parameter efficiency and local-global feature modeling. However, the computational needs and resources required for such models make them less suitable for deployment in under-resourced environments. Another notable approach, the Two-Track UNet \cite{HTTU}, utilizes dual convolutional tracks to address class imbalances, yet its reliance on specific kernel configurations reduces adaptability to heterogeneous datasets. Finally, H-DenseUNet \cite{HDENSENET} fuses 2D and 3D DenseUNets to handle intra- and inter-slice features, achieving strong performance on liver and tumor segmentation tasks. Despite its effectiveness, the model’s dependence on large-scale public datasets and computational resources creates challenges for real-world clinical deployment.
\paragraph{}
This study introduces a hybrid U-Net architecture that integrates a \textbf{Transformer Bottleneck} and \textbf{Attention Gates}, tailored specifically to handle the heterogeneity of local clinical datasets. The architecture builds upon the well-established U-Net framework by enhancing its encoder-decoder structure with modified components from previously used architectures in literature, leveraging the strengths of both convolutional and transformer-based representations. 

\begin{figure}
    \centering
    \includegraphics[width=1\linewidth]{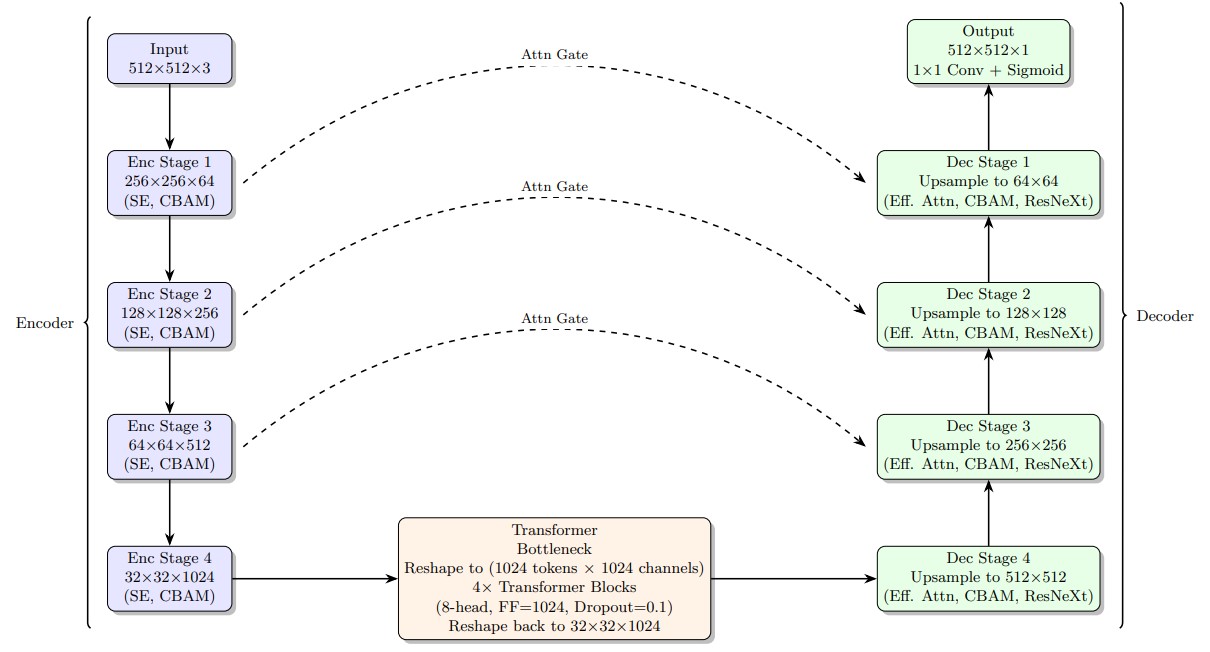}
    \caption{Overview of the proposed architecture. The Encoder (blue) uses a ResNet50-based U-Net with SE and CBAM modules to extract hierarchical features. The Transformer Bottleneck (orange) processes these features via four transformer blocks. The Decoder (green) upsamples and fuses encoder features through skip connections (labeled “Attn Gate”) refined by efficient attention, CBAM, and ResNeXt modules, yielding the final segmentation map. }
    \label{fig:architecture}
\end{figure}

Our proposed hybrid architecture (see Figure \ref{fig:architecture}) builds upon the canonical U-Net framework to achieve precise segmentation of clinical images while addressing the limitations of resource-constrained training environments. In the Encoder, we utilize a ResNet-50\cite{resnet50} backbone from the \textbf{$segmentation\_models$}\cite{yakubovskiy2019segmentation} library, pre-trained on ImageNet\cite{imagenet}. This design choice is motivated not only by ResNet-50’s\cite{resnet50} proven ability to extract robust hierarchical features from $512 \times512\times3$ input images but also by the practical necessity to reduce training time and computational cost on Kaggle GPUs, which impose strict limits on available resources. The encoder produces feature maps at four distinct stages: Stage 1 yields a $256\times256\times64$ tensor capturing low-level details (edges and textures), Stage 2 outputs a $128\times128\times256$ tensor representing medium-level features, Stage 3 produces a $64\times64\times512$ tensor encapsulating high-level information, and Stage 4 delivers a $32\times32\times1024$ tensor that abstracts semantic content. To further enhance these representations, each encoder block is augmented with Squeeze-and-Excitation (SE)\cite{hu2018squeeze} and Convolutional Block Attention Modules (CBAM)\cite{woo2018cbam}. The SE block, for example, recalibrates the input $X \in \mathbb{R}^{H\times W\times C}$ by first computing a channel descriptor via global average pooling,$z = GAP(X)$ then passing it through two FC layers with a reduction ratio r (typically 16): $s = \sigma(W_2 \times ReLU(W_1 z) $ and finally re-weighting the input by performing element-wise multiplication $X^{'} = X\odot s$. This mechanism enhances the network’s sensitivity to clinically important features, such as subtle tumor boundaries.

At the network’s core, the Transformer Bottleneck is employed to capture long-range dependencies that conventional convolutional layers cannot. Here, the deepest encoder output is a $32 \times32\times1024$ tensor, that is reshaped into a sequence of 1024 tokens (i.e., $1024\times1024$) and processed through a series of four transformer blocks. Each block applies multi-head self-attention\cite{attention} (as described in the previously provided Equation \ref{eq1} and a feedforward network with 1024 hidden units, along with dropout (0.1) and layer normalization. This design significantly broadens the receptive field and integrates global contextual information, which is essential for delineating diffuse and irregular tumor regions.

In the Decoder, the network gradually upsamples the feature maps to restore the original $512 \times512$ resolution. At every upsampling stage, skip connections fuse high-resolution encoder features into the decoder output. Before fusion, the skip connection features are refined using CBAM to recalibrate channel and spatial responses. Moreover, an efficient attention\cite{shen2021efficient} mechanism is applied within the decoder to dynamically emphasize the most informative features. In this module, the input tensor is first reduced in channel dimension using a $1 \times1$ convolution, followed by an additive branch and a ReLU activation; a subsequent $1 \times1$ convolution, followed by a sigmoid activation, yields attention coefficients $\psi$ that modulate the original feature map as $Y =  X \odot \psi$, thereby filtering out irrelevant information. To further refine these features, each decoder block incorporates additional SE blocks and ResNeXt\cite{resNeXt} modules. The ResNeXt block leverages grouped convolutions to capture multi-scale spatial patterns efficiently. It first reduces channel dimensions via a $1 \times1$ convolution, then processes spatial information using a depthwise $3 \times3$ convolution, and finally expands the channel dimension with another $1 \times1$ convolution. A parallel shortcut branch ensures that the block benefits from residual learning, and a final ReLU activation completes the operation.

Finally, the refined features are passed through a $1 \times1$ convolution with sigmoid activation, that generates a binary segmentation map. By combining the data efficiency and strong inductive bias of CNNs with the global context modeling capabilities of transformers, our hybrid architecture achieves superior segmentation performance even when trained on limited annotated data, which is a key advantage in clinical settings. The use of pre-trained ImageNet weights in the encoder not only accelerates convergence but also mitigates the computational burden on Kaggle GPUs, making the entire framework both efficient and precise.

Figure \ref{fig:architecture} provides an overview of our hybrid U-Net architecture, which is specifically tailored for segmentation of heterogeneous clinical images. In the Encoder (left, shown in blue), we employ a ResNet50-based U-Net that leverages convolutional layers to extract hierarchical features from 512×512 RGB images. Each encoder block is enhanced with Squeeze-and-Excitation (SE) and Convolutional Block Attention Modules (CBAM), which recalibrate channel and spatial responses to accentuate critical features such as tumor boundaries. These modifications are crucial in our context, as they improve the representation of subtle, clinically relevant structures in data where annotations are scarce.

At the core of our architecture, the Transformer Bottleneck (depicted in orange in Figure \ref{fig:architecture}) addresses the inherent limitation of CNNs in modeling long-range dependencies. Here, the deepest encoder feature map is reshaped into a sequence of tokens, which are processed through four transformer blocks using 8-head self-attention, feed-forward layers, dropout, and residual connections. This design not only broadens the receptive field but also enables the model to integrate global contextual information—a key requirement for accurately delineating diffuse and irregular tumor regions.

In the Decoder (right, shown in green), the network gradually upsamples the feature maps back into an image of original resolution. Skip connections, as indicated by the dashed arrows labeled “Attn Gate” in Figure Figure \ref{fig:architecture}, fuse high-resolution encoder features with the decoder output. Prior to fusion, each skip connection is refined using CBAM, and then the decoder applies an efficient attention mechanism to selectively emphasize the most informative features. Furthermore, the decoder interleaves additional SE blocks and ResNeXt blocks; the latter leverages grouped convolutions\cite{resNeXt} to capture multi-scale patterns efficiently, ensuring that both local details and global context are preserved. This comprehensive integration of modules enables our architecture to balance the data efficiency of convolutional networks\cite{unet} with the long-range image dependency modeling of transformers\cite{chen2021transunet}, thus achieving robust segmentation performance even on limited clinical datasets.

\subsection{The Training of Hybrid Architecture}

The proposed hybrid architecture was trained end-to-end using a composite loss function, meticulously designed to leverage the strengths of both region-based and pixel-wise losses. This approach ensures robust segmentation performance, particularly when working with resource-constrained clinical datasets characterized by limited annotated samples. 

The \textbf{loss function} combines Binary Cross-Entropy (BCE) loss and soft Dice loss in a weighted formulation. The soft Dice loss, which mitigates class imbalance by focusing on region overlap, is expressed as: 
\begin{equation}
    DiceLoss = 1 - \frac{2\sum_i(\hat{y}_i\cdot y_i)+\epsilon}{\sum_i(\hat{y}_i) + \sum_i(y_i) + \epsilon}
\end{equation}
where $\hat{y}_i$ and $y_i$ denote the predicted and ground-truth probabilities for pixel $i$, and $\epsilon$ is a small constant added to prevent division by zero. 
\begin{equation}
    \mathcal{L}_{overall} = \mathcal{L}_{BCE} + \lambda\cdot\mathcal{L}_{Dice}
\end{equation}
where $\lambda$ is a hyperparameter empirically optimized to balance the contributions of the two losses. This hybrid formulation has demonstrated significant effectiveness in balancing global context with fine-grained segmentation accuracy in medical image segmentation tasks \cite{Milletari}.

\subsubsection{Evaluation Metrics}
The segmentation performance is assessed based on some evaluation metrics, we employed several metrics commonly used in medical imaging studies \cite{dataset1, HTTU, HDENSENET}. These metrics include \textbf{Dice Score}, which is a measure of the degree of overlap between predicted and ground truth image, similar to \textbf{Intersection over Union(IoU)}, which quantifies the ratio of the intersection to the union of predicted and actual regions. There are several metrics that are traditionally used in machine learning  literature, namely \textbf{Precision}, which is basically a metric that quantifies the positive predictive value of the model, and \textbf{Recall} that measures the model’s sensitivity to detecting positive regions. 
\paragraph{}
These metrics provide a multi-dimensional evaluation of the model, ensuring that it achieves a balance between regional accuracy, global overlap, and fine-grained delineation.
\paragraph{}
To optimize computational efficiency, we initialized the encoder of the hybrid model with pre-trained weights derived from ImageNet\cite{imagenet}. This transfer learning approach reduced convergence time and enhanced the model’s ability to extract domain-relevant features from limited annotated data, as previously demonstrated in medical image segmentation tasks \cite{unet++}. The training process was conducted on GPUs provided by Kaggle, imposing constraints on computational resources. Nevertheless, through the use of pretraining and an empirically fine-tuned loss function, the currently employed architecture achieved a notably high level of accuracy while also being quite computationally efficient. The combined strategies ensure that the model is robust, efficient, and capable of performing effectively on local clinical datasets with inherent variability.

 \section{Results and Discussion}

Our proposed hybrid architecture was trained end-to-end on a clinically sourced MRI dataset comprising 6,080 images, expanded from an initial set of approximately 1,000 images through extensive data augmentation. This approach contrasts with studies utilizing larger, publicly available datasets such as BraTS, highlighting our model's efficacy in handling limited, heterogeneous clinical data. Training was conducted on Kaggle's GPU infrastructure, which imposes a 12-hour runtime limitation. To navigate this constraint, we implemented a robust checkpointing mechanism, enabling seamless resumption of training sessions and ensuring uninterrupted model development. 
\begin{figure}[htbp]
    \centering
    \begin{subfigure}{0.32\linewidth}
        \centering
        \includegraphics[width=\linewidth]{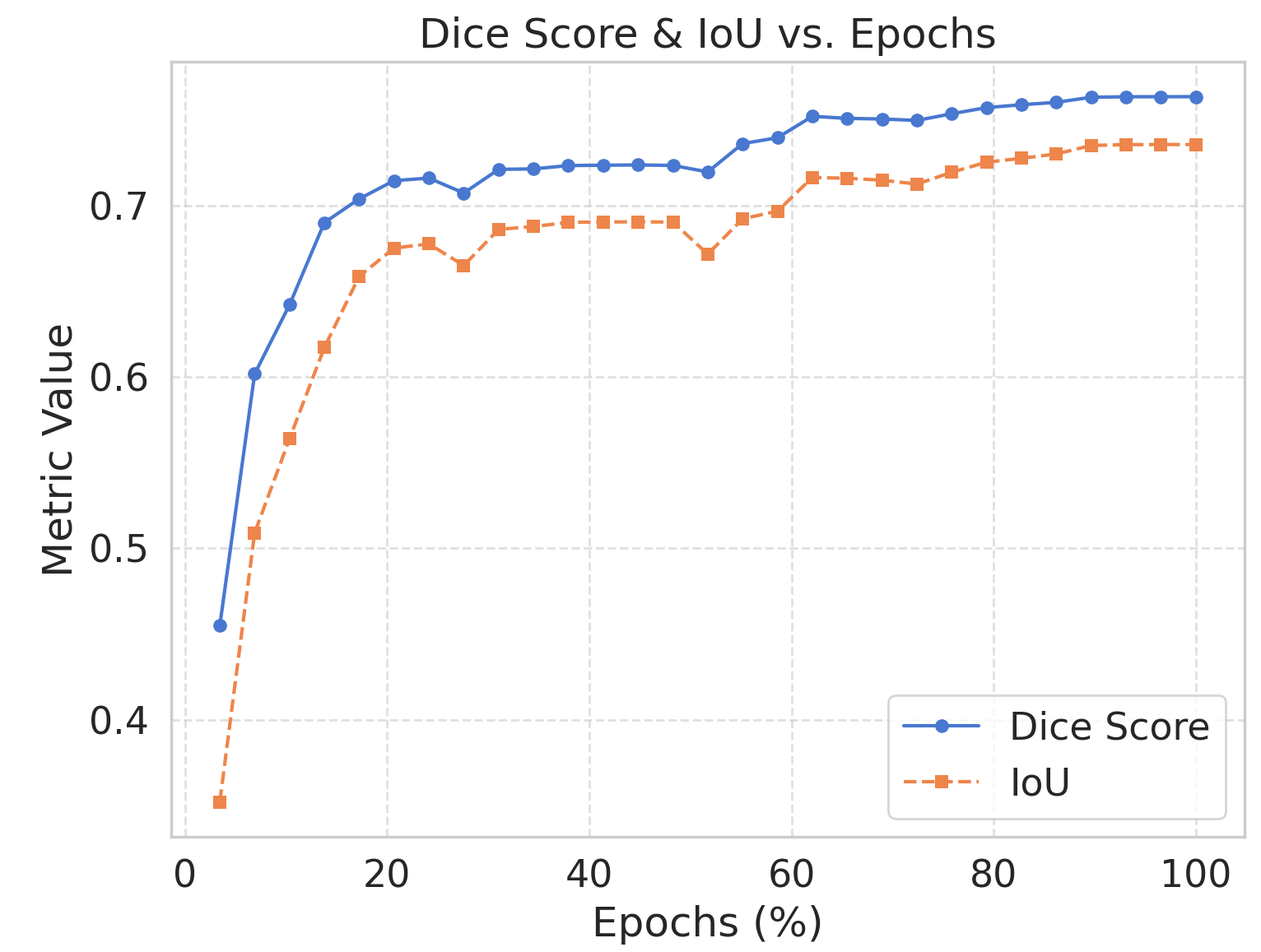}
        \caption{Dice Score and IoU}
        \label{fig:DiceScoreandIOU}
    \end{subfigure}
    \hfill
    \begin{subfigure}{0.32\linewidth}
        \centering
        \includegraphics[width=\linewidth]{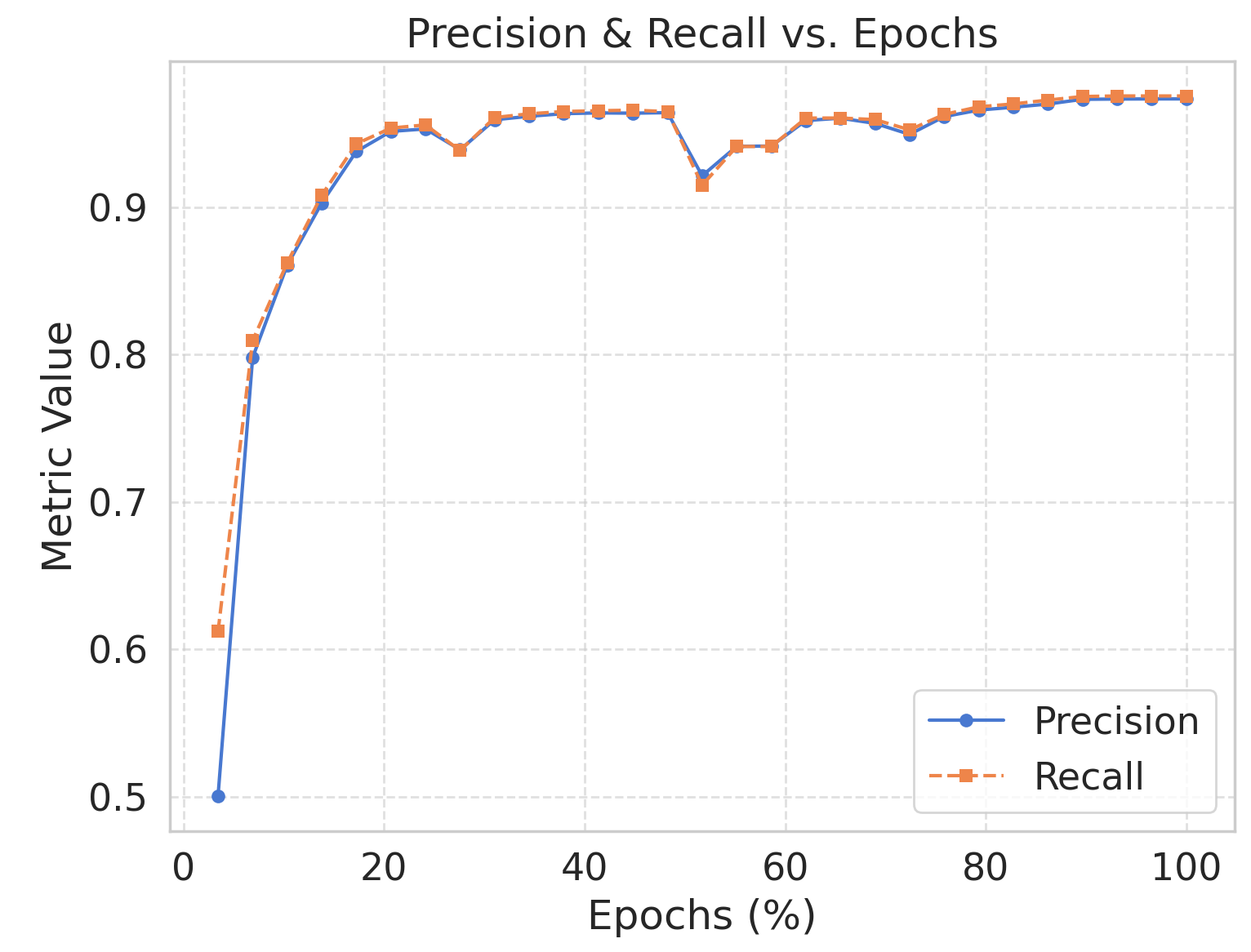}
        \caption{Precision and Recall}
        \label{fig:precisionandrecall}
    \end{subfigure}
    \hfill
    \begin{subfigure}{0.32\linewidth}
        \centering
        \includegraphics[width=\linewidth]{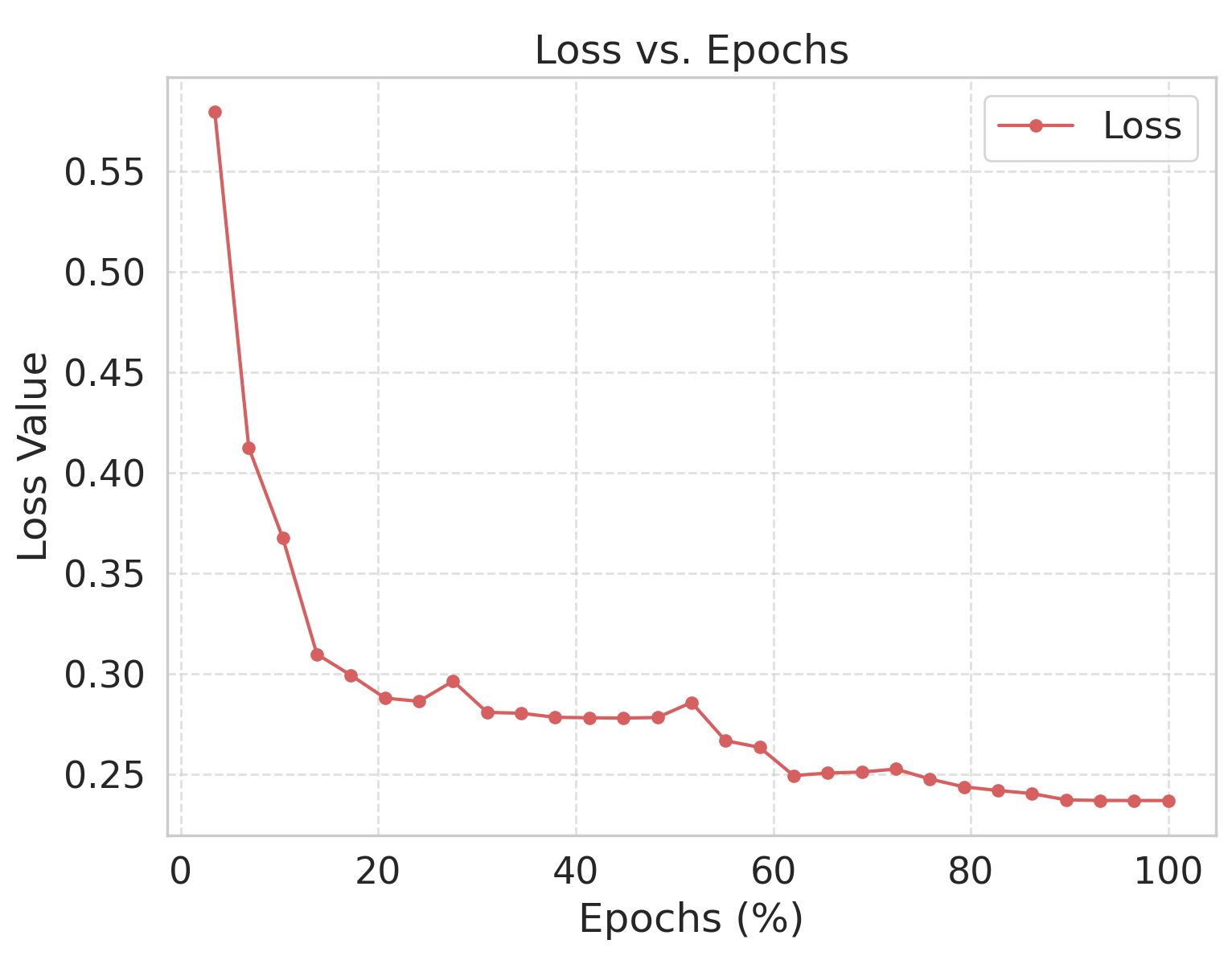}
        \caption{Loss}
        \label{fig:Loss}
    \end{subfigure}
    \caption{Training performance metrics over epochs.}
    \label{fig:overall_performance}
\end{figure}

As depicted in Figure \ref{fig:DiceScoreandIOU}, the model exhibited consistent convergence over 29 epochs with a batch size of 8. The initial Dice Score and IoU were 0.4548 and 0.3519, respectively. By epoch 29, these metrics improved to 0.7636 and 0.7357, respectively. The overall loss decreased from 0.5798 to 0.2368, as shown in Figure \ref{fig:precisionandrecall}, while precision and recall increased to 0.9736 and 0.9756, respectively. The hybrid loss also steadily decreased from 0.5798 to 0.2368 as depicted in Figure \ref{fig:Loss}. These trends underscore the model's learning efficiency and the effectiveness of the composite loss function in balancing regional overlap with pixel-level accuracy. 

\begin{table}[htbp]
  \centering
  \renewcommand{\arraystretch}{1.3}
  \resizebox{0.9\linewidth}{!}{%
    \begin{tabular}{>{\centering\arraybackslash}p{2.5cm} >{\centering\arraybackslash}p{3cm} >{\centering\arraybackslash}p{2.5cm} >{\centering\arraybackslash}p{2.5cm}@{}}
      \toprule
       \textbf{Model} & \textbf{Dataset Used} & \textbf{Number of Slices} & \textbf{Dice Score / IoU} \\ \midrule
       HTTU-Net & BraTS 2018 & $\sim171,000$ & 0.865 / 0.745 \\
       H-DenseUNet & LiTS 2017   & 58,638     & 0.882 / 0.792 \\
       MM-BiFPN   & BraTS 2018 \& BraTS 2020 & $\sim150,000+$ & 0.8107 / 0.7031 \\
       Hybrid Model & Local Clinical Dataset & 6,080      & 0.7636 / 0.7357 \\ \bottomrule
    \end{tabular}%
  }
  \caption{Comparison of our final model with other models in the literature.}
  \label{tab:lit_comparison}
\end{table}

\begin{table}[htbp]
  \centering
  \begin{tabular}{@{}lcccc@{}}
    \toprule
    \textbf{Model} & \textbf{Dice Score} & \textbf{IoU} & \textbf{Precision} & \textbf{Recall} \\ \midrule
    U-Net (Baseline)                  & 0.7190 & 0.6818 & 0.9500 & 0.9400 \\
    U-Net with Transformer Bottleneck & 0.7286 & 0.7018 & 0.9600 & 0.9500 \\
    \textbf{Final Hybrid Model}       & 0.7636 & 0.7357 & 0.9736 & 0.9756 \\ \bottomrule
  \end{tabular}
  \caption{Performance comparison of our model variants.}
  \label{tab:model_variants}
\end{table}

A comparative analysis against leading architectures is shown in Table  \ref{tab:lit_comparison}. Aboelenein et al. (2020)\cite{HTTU} proposed HTTU Net and reported a Dice Score of 0.865 and an IoU of 0.745 on the BraTS 2018 dataset, which comprises roughly 171 000 slices. Likewise, Li et al. (2018)\cite{HDENSENET} developed H DenseUNet and achieved a Dice Score of 0.882 and an IoU of 0.792 on the LiTS 2017 dataset of 58 638 slices. Syazwany et al. (2021)\cite{MMBiFPN} introduced MM BiFPN, obtaining a Dice Score of 0.8107 and an IoU of 0.7031 on the BraTS 2018 and 2020 datasets, which include about 171 000 and 221 400 slices respectively. By comparison, our final hybrid model integrates the Transformer Bottleneck, efficient attention\cite{shen2021efficient} mechanisms, Squeeze and Excitation (SE)\cite{hu2018squeeze} blocks, Convolutional Block Attention Module (CBAM)\cite{woo2018cbam}, and ResNeXt modules, achieving a Dice Score of 0.7636 and an IoU of 0.7357. It was trained on a local clinical dataset of only 6 080 slices. Considering the challenges of working with a small heterogeneous clinical dataset under limited computing resources, these results highlight the robustness and effectiveness of the proposed architecture.

Table  \ref{tab:model_variants} presents a comparison with baseline models. The standard U Net achieved a Dice Score of 0.7190 and an IoU of 0.6818. Incorporating only a Transformer Bottleneck raised these values to a Dice Score of 0.7286 and an IoU of 0.7018. Our final hybrid model integrates the Transformer Bottleneck, efficient attention mechanisms, Squeeze and Excitation (SE) blocks, Convolutional Block Attention Module (CBAM), and ResNeXt modules, achieving a Dice Score of 0.7636 and an IoU of 0.7357. Considering the challenges of training on a small heterogeneous clinical dataset under computational constraints, these findings underscore the robustness and effectiveness of the proposed design.

\subsection{Discussion}
The integration of transformer-based global context with localized attention mechanisms in a ResNet50-based U-Net framework significantly enhances segmentation performance. The Transformer Bottleneck broadens the receptive field by reshaping deep encoder features into tokens and processing them through multi-head self-attention, capturing long-range dependencies. This global context is complemented by efficient attention mechanisms within the decoder, which dynamically emphasize the most informative features. The inclusion of SE blocks and ResNeXt modules facilitates robust multi-scale feature fusion, contributing to the high Dice Score and IoU achieved by the model.

The use of pre-trained ImageNet weights in the encoder was instrumental in reducing convergence time and computational costs, a strategy similarly employed in models like H-DenseUNet for liver and tumor segmentation \cite{HDENSENET}. This transfer learning approach allowed us to leverage large-scale feature representations, which is particularly advantageous given the limited GPU resources and training time available on Kaggle. The checkpointing system further ensured that the training process could be effectively managed despite these constraints.

In summary, our experimental findings validate the efficacy of the proposed hybrid architecture, demonstrating its potential for precise medical image segmentation on relatively small, heterogeneous clinical datasets. The improvements over the baseline U-Net and the U-Net with only a Transformer Bottleneck highlight the value of integrating efficient attention mechanisms and multi-scale feature fusion strategies under constrained computational environments.

 \section{Conclusion and Future Work}
 In this paper, we introduced a novel hybrid architecture for tumor segmentation in MRI images that integrates a ResNet50-based U-Net with a Transformer Bottleneck, efficient attention mechanisms, SE and CBAM modules, and ResNeXt blocks. Our approach leverages the strengths of both convolutional networks and transformers, resulting in enhanced local feature extraction and global context modeling. Trained on a limited clinical dataset using Kaggle GPUs with a batch size fixed at 8 the model achieved a Dice Score of 0.7636 and an IoU of 0.7357, demonstrating competitive performance despite constrained resources.

The use of pre-trained ImageNet weights in the encoder not only accelerated convergence but also reduced computational demands, making the model efficient and robust for clinical applications. Our extensive evaluation confirms that the proposed method effectively balances pixel-wise accuracy with regional overlap, and that the integration of advanced attention modules significantly improves segmentation outcomes in heterogeneous clinical data.

For future work, we plan to expand our collaboration with the local hospital to acquire a larger and more diverse dataset, including a wider range of imaging modalities. With access to enhanced computational resources at our university, we aim to explore training with larger batch sizes, which may further stabilize training and boost performance. Additionally, we will investigate the extension of our method to 3D segmentation to better capture volumetric information, and consider integrating additional self-supervised learning strategies to further refine feature representations.

Overall, our study demonstrates that a carefully designed hybrid architecture can achieve precise segmentation on limited clinical datasets while remaining computationally efficient—a promising step toward practical clinical deployment.

\printbibliography

\end{document}